
\input lanlmac

\def\d{\partial}
\def\pder#1{{\partial\over\partial{#1}}}

\def\o{\over}
\def\hf{{1 \over 2}}
\def\bra#1{\left\langle #1 \right|}
\def\ket#1{\left| #1 \right\rangle}

\def\sitarel#1#2{\mathrel{\mathop{\kern0pt #1}\limits_{#2}}}
\def\inbar{\,\vrule height1.5ex width.4pt depth0pt}
\def\IC{\relax\hbox{$\inbar\kern-.3em{\rm C}$}}
\def\IR{\relax{\rm I\kern-.18em R}}
\def\IP{\relax{\rm I\kern-.18em P}}

\def\a{\alpha}
\def\A{\Psi}
\def\dB{\delta{B}}
\def\l{l_s}
\def\QB{Q_{\rm B}}
\def\hQB{\hat{Q}_{\rm B}}
\def\ro{{\rm o}}
\def\rc{{\rm c}}
\def\hX{\hat{X}}
\def\hVo#1{\left\langle\right.\kern-2pt \hat{V}_3^{\rm o} (#1) 
\kern-2pt\left.\right|}
\def\hVfo#1{\left\langle\right.\kern-2pt \hat{V}_4^{\rm o}(#1)
\kern-2pt\left.\right|}
\def\hVc#1{\left\langle\right.\kern-2pt \hat{V}_3^{\rm c}(#1)\kern-2pt
\left.\right|}
\def\hU#1{\left\langle\right.\kern-2pt \hat{U}(#1)\kern-2pt\left.\right|}
\def\hUomg#1{\left\langle\right.\kern-2pt \hat{U}_\Omega(#1)
\kern-2pt\left.\right|}
\def\hvo#1{\left\langle\right.\kern-2pt \hat{v}_3^{\rm o}(#1)
\kern-2pt\left.\right|}
\def\hvfo#1{\left\langle\right.\kern-2pt \hat{v}_4^{\rm o}(#1)
\kern-2pt\left.\right|}
\def\hvc#1{\left\langle\right.\kern-2pt \hat{v}_3^{\rm c}(#1)
\kern-2pt\left.\right|}
\def\hu#1{\left\langle\right.\kern-2pt \hat{u}(#1)\kern-2pt\left.\right|}
\def\huomg#1{\left\langle\right.\kern-2pt \hat{u}_\Omega(#1)
\kern-2pt\left.\right|}
\def\Vo#1{\left\langle V_3^\ro(#1)\right|}
\def\Vfo#1{\left\langle V_4^{\rm o}(#1)\right|}
\def\Vc#1{\left\langle V_3^{\rm c}(#1)\right|}
\def\U#1{\left\langle U(#1)\right|}
\def\Uomg#1{\left\langle U_\Omega(#1)\right|}
\def\vo#1{\left\langle v_3^{\rm o}(#1)\right|}
\def\vfo#1{\left\langle v_4^{\rm o}(#1)\right|}
\def\vc#1{\left\langle v_3^{\rm c}(#1)\right|}
\def\u#1{\left\langle u(#1)\right|}
\def\uomg#1{\left\langle u_\Omega(#1)\right|}

\def\np#1#2#3{{ Nucl. Phys.} {\bf B#1}, #2 (#3)}

\def\pln#1#2#3{{Phys. Lett. } {\bf B#1}, #2 (#3)}
\def\plo#1#2#3{{ Phys. Lett.} {\bf #1B}, #2 (#3)}
\def\pr#1#2#3{{ Phys. Rev.} {\bf D#1}, #2 (#3)}
\def\prl#1#2#3{{ Phys. Rev. Lett.} {\bf #1}, #2 (#3) }

\def\ann#1#2#3{{ Ann. Phys.} {\bf #1}, #2 (#3)}
\def\ptp#1#2#3{{ Prog. Theor. Phys.} {\bf #1}, #2 (#3)}

\def\mpl#1#2#3{{ Mod. Phys. Lett.} {\bf A#1}, #2 (#3)}
\def\jhep#1#2#3{{JHEP} {\bf #1}, #2 (#3)}

\def\hpt#1{{\tt hep-th/#1}}

\lref\CDS{
A.~Connes, M. R.~Douglas, and A.~Schwarz, 
``Noncommutative Geometry and Matrix Theory: Compactification on Tori,''
\jhep{02}{003}{1998}, \hpt{9711162}.
}
\lref\DH{
M. R.~Douglas and C.~Hull,
``D-Branes and the Noncommutative Torus,''
\jhep{02}{008}{1998}, \hpt{9711165}\semi
Y.-K. E. Cheung and M. Krogh,
``Noncommutative Geometry from 0-branes in a Background B Field,''
\np{528}{185}{1998}, \hpt{9803031}\semi
T. Kawano and K. Okuyama,
``Matrix Theory on Noncommutative Torus,''
\pln{433}{29}{1998}, \hpt{9803044}.
}
\lref\Schom{
V. Schomerus,
``D-branes and Deformation Quantization,''
\jhep{9906}{030}{1999}, \hpt{9903205}.
}
\lref\SW{
N.~Seiberg and E.~Witten, 
``String Theory and Noncommutative Geometry,''
\jhep{9909}{032}{1999}, \hpt{9908142}.
}
\lref\AsaKis{
T.~Asakawa and I.~Kishimoto,
``Comments on Gauge Equivalence in Noncommutative Geometry,''
\jhep{9911}{024}{1999}, \hpt{9909139}.
}
\lref\SWrel{
L.~Cornalba, 
``D-brane Physics and Noncommutative Yang-Mills Theory,''
\hpt{9909081}\semi
N.~Ishibashi, 
``A Relation between Commutative and Noncommutative Descriptions of D-branes,''
\hpt{9909176}\semi
K.~Okuyama, 
``A Path Integral Representation of the Map between
Commutative and Noncommutative Gauge Fields,''
\hpt{9910138}.
}
\lref\WSFT{
E.~Witten, 
``Noncommutative Geometry and String Field Theory,''
\np{268}{253}{1986}.
}
\lref\HIKKO{
H.~Hata, K.~Itoh, T.~Kugo, H.~Kunitomo, and K.~Ogawa,
``Covariant String Field Theory,''
\pr{34}{2360}{1986}.
}
\lref\Sen{
A.~Sen,
``Open String Field Theory in Nontrivial Background Field (I): 
Gauge Invariant Action,''
\np{334}{350}{1990}.
}
\lref\MS{
T. R. Morris and B. Spence,
``Background Independent String Field Theory,''
\np{316}{113}{1989}.
}
\lref\HLRS{
G. T.~Horowitz, J.~Lykken, R.~Rohm, and A.~Strominger,
``Purely Cubic Action for String Field Theory,''
\prl{57}{283}{1986}.
}
\lref\HIKKOpre{
H.~Hata, K.~Itoh, T.~Kugo, H.~Kunitomo, and K.~Ogawa,
``Pregeometrical String Field Theory: Creation of Space-Time and Motion,''
\plo{175}{138}{1986}.
}
\lref\ChuHo{
C.-S.~Chu and P.-M.~Ho, 
``Noncommutative Open String and $D$-Brane,''
\np{550}{151}{1999}, \hpt{9812219};
``Constrained Quantization of Open String in Background $B$ Field and 
Noncommutative $D$-Brane,'' 
\hpt{9906192}.
}
\lref\Jabbari{
F. Ardalan, H. Arfaei, and M. M. Sheikh-Jabbari, 
``Dirac Quantization of Open Strings and Noncommutativity in Branes,''
\hpt{9906161}.
}
\lref\HS{
G. T.~Horowitz and A.~Strominger,
``Translations as Inner Derivations and Associativity Anomaly in 
Open String Field Theory,'' 
\pln{185}{45}{1987}.
}
\lref\LPP{
A.~LeClair, M.~Peskin, and C. R.~Preitschopf,
``String Field Theory on the Conformal Plane (I),''
\np{317}{411}{1989}.
}
\lref\GJ{
D. J.~Gross and A.~Jevicki,
``Operator Formulation of Interacting String Field Theory (I),''
\np{283}{1}{1987};
``Operator Formulation of Interacting String Field Theory (II),''
\np{287}{225}{1987};
``Operator Formulation of Interacting String Field Theory (III), 
NSR Superstring,''
\np{293}{29}{1987}.
}
\lref\Yoneya{
T.~Yoneya,
``String Coupling Constant and Dilaton Vacuum Expectation Value in String 
Field Theory,''
\pln{197}{76}{1987}.
}
\lref\CST{
E.~Cremmer, A.~Schwimmer, and C.~Thorn,
``The Vertex Function in Witten's Formulation of String Field Theory,''
\plo{179}{57}{1986}.
}
\lref\FS{
M.~Fisk and M.~Srendnicki,
``Magnetic String Fields,''
\np{313}{308}{1989}.
}
\lref\WittenSSFT{
E.~Witten,
``Interacting Field Theory of Open Superstrings,''
\np{276}{291}{1986}.
}
\lref\Suehiro{
K.~Suehiro,
``Operator Expression of Witten's Superstring Vertex,''
\np{296}{333}{1988}.
}
\lref\Wendt{
C.~Wendt,
``Scattering Amplitudes and Contact Interactions in
Witten's Superstring Field Theory,''
\np{314}{209}{1989}.
}
\lref\KugoZwie{
T.~Kugo and B.~Zwiebach,
``Target Space Duality as a Symmetry of String Field Theory,''
\ptp{87}{801}{1992}.
}
\lref\Berkovits{
N.~Berkovits and C. T.~Echevarria,
``Four-Point Amplitude from Open Superstring Field Theory,''
\hpt{99012120}\semi
N.~Berkovits,
``A New Approach to Superstring Field Theory,''
\hpt{99012121}.
}
\lref\Sugino{
F.~Sugino,
``Witten's Open String Field Theory in Constant $B$-field Background,''
\jhep{0003}{017}{2000}, \hpt{9912254}.
}
\lref\Wopenclosed{
B.~Zwiebach, 
``Oriented Open-Closed String Field Theory Revisited,''
\ann{267}{193}{1998}, \hpt{9705241}; 
``Quantum Open String Theory with Manifest Closed String Factorization,''
\pln{256}{22}{1991};
``Interpolating String Field Theories,'' 
\mpl{7}{1079}{1992}, \hpt{9202015}.
}
\lref\KT{
T.~Kugo and  T.~Takahashi,
``Unoriented Open-Closed String Field Theory,''
\ptp{99}{649}{1998}, \hpt{9711100}.
}
\lref\AKTa{
T.~Asakawa, T.~Kugo, and  T.~Takahashi,
``BRS Invariance of Unoriented Open-Closed String Field Theory,''
\ptp{100}{831}{1999}, \hpt{9807066}.
}
\lref\AKTb{
T.~Asakawa, T.~Kugo, and  T.~Takahashi,
``One-Loop Tachyon Amplitude in Unoriented Open-Closed String Field Theory,''
\ptp{102}{427}{1999}, \hpt{9905043}
}
\lref\AKTc{
T.~Asakawa, T.~Kugo, and  T.~Takahashi,
``On the Generalized Gluing and Resmoothing Theorem,''
\ptp{437}{100}{1998}
}
\lref\Polchn{
J.~Polchinski, 
``Dirichlet Branes and Ramond-Ramond Charges,''
\prl{75}{4724}{1995}, \hpt{9510017}.
}
\lref\DBI{
R. G.~Leigh,
``Dirac-Born-Infeld Action from Dirichlet Sigma Model,''
\mpl{4}{2726}{1989}.
}
\lref\FS{
W.~Fischler and L.~Susskind,
``Dilaton Tadpoles, String Condensates and Scale Invariance,''
\pln{171B}{383}{1986};
``Dilaton Tadpoles, String Condensates and Scale Invariance 2,''
\pln{173B}{262}{1987}.
}
\lref\KugoSuppl{
T.~Kugo,
``Unoriented Open-Closed String Field Theory,''
Prog. Theor. Phys. Suppl. {\bf 134}, 137, (1999).
}
\lref\hiko{
T.~Kawano and T.~Takahashi,
``Open String Field Theory on Noncommutative Space,''
\hpt{9912274}.
}
\lref\HIKKOclosed{
H.~Hata, K.~Itoh, T.~Kugo, H.~Kunitomo, and K.~Ogawa,
``Covariant String Field Theory. II,''
\pr{35}{1318}{1987}.
}
\lref\HN{
H.~Hata and M. M.~Nojiri,
``New Symmetry in Covariant Open-String Field Theory,''
\pr{36}{1193}{1987}.
}
\lref\Mandelstam{
S.~Mandelstam,
``Interacting-String Picture of Dual Resonance Models,''
\np{64}{205}{1973}.
}
\lref\NeveuWest{
A. Neveu and P. West, 
``The Interaction Gauge Covariant Bosonic String,''
\pln{169}{192}{1986}.
}
\lref\KakuKikkawa{
M. Kaku and K. Kikkawa, 
``The Field Theory of Relativistic Strings, PT. 1: Trees,''
\pr{10}{1110}{1974}.
}
\lref\Strominger{
A. Strominger, 
``Closed Strings in Open String Field Theory,''
\prl{58}{629}{1987}.
}
\lref\AGGMV{
L. Alvarez-Gaum\'e, C. Gomez, G. Moore, and C. Vafa, 
``Strings in the Operator Formalism,''
\np{303}{455}{1988}
}
\lref\GiddMarti{
S. B. Giddings and E. Martinec, 
``Conformal Geometry and String Field Theory,''
\np{278}{91}{1986}
}


\Title{                                \vbox{\hbox{UT-880}
                                             \hbox{\tt hep-th/0005080}} }
{\vbox{\hbox{\centerline{
                    Open-Closed String Field Theory 
}}
\vskip .05in
\hbox{\centerline{
                      in the Background $B$-Field
}}}}

\centerline{
                            Teruhiko Kawano 
}
\centerline{\sl
                      Department of Physics, University of Tokyo
}
\centerline{\sl
                            Hongo, Tokyo 113-0033, Japan
}
\centerline{\tt
                         kawano@hep-th.phys.s.u-tokyo.ac.jp
}
\centerline{and}
\centerline{
                             Tomohiko Takahashi 
}
\centerline{\sl
                      Department of Physics, Nara Women's University
}
\centerline{\sl
                            Nara 630-8506, Japan
}
\centerline{\tt
                             tomo@phys.nara-wu.ac.jp
}

\vskip 0.5in

In this paper, we study open-closed string field theory 
in the background B-field 
in the so-called $\alpha=p^{+}$ formulation.
The string field theory in the infrared gives noncommutative gauge 
theory in the open string sector. 
Since this theory includes closed string fields as dynamical variables, 
we can obtain another string field theory in the same background through the 
condensation of closed string fields, whose low-energy effective action 
does not show the noncommutativity of spacetime. 
Although we have two string field theories in the same background, 
we show that these theories are equivalent. 
In fact, we give the unitary transformation from string fields in one of 
them to string fields in the other.

\bigskip
\Date{May, 2000}


\newsec{Introduction}

In our previous paper \hiko, we studied Witten's open string field theory 
\WSFT\ in the background $B$-field \refs{\hiko,\Sugino}. In the resulting 
string field theory, the kinetic term has an ordinary form, except that 
the metric in the BRS charge of the kinetic term is the open string metric 
$G_{ij}$, not the closed string metric $g_{ij}$. The open string metric 
$G_{ij}$ is related to the closed string metric $g_{ij}$ by 
$G_{ij}=g_{ij}-b_{ik}g^{kl}b_{lj}$, where $b_{ij}=(2\pi\alpha')B_{ij}$; 
$B_{ij}$ is the background $B$-field. The interaction term is also affected by 
the background $B$-field.  The interaction term of Witten's open string field 
theory is only the three-string vertex 
$
{}_{321}\left\langle V_3 \right|
\left|\Psi\right\rangle_1\left|\Psi\right\rangle_2\left|\Psi\right\rangle_3
$. In the presence of the $B$-field, the three-string vertex 
$\left\langle V_3 \right|$ becomes 
$\left\langle V_3 \right|\exp[-(i/4)\sum_{r<s}^3\theta^{ij}p_i^{(r)}p_j^{(s)}]$. 
These results are in agreement with the results in \SW. 
Therefore, we can expect that the low-energy physics is effectively described 
by noncommutative gauge theory. 

The novelty of the papers \refs{\hiko,\Sugino} is that 
the noncommutative factor 
$e^{[-(i/4)\sum\theta^{ij}p_i^{(r)}p_j^{(s)}]}$ on the three-string 
vertex $\left\langle V_3 \right|$ can be expressed as 
$\prod^{3}_{r=1}U^{(r)}$, where $U^{(r)}=e^{M^{(r)}}$ is 
a unitary operator of the $r$-th 
string. Therefore, it is very tempting to transform open string fields $\Psi$ by 
$\Psi\rightarrow U^{-1}\Psi$ to remove the noncommutative factor 
$\exp[-(i/4)\sum_{r<s}^3\theta^{ij}p_i^{(r)}p_j^{(s)}]$ from the three-string 
vertex, and to expect that the low-energy dynamics is described by an ordinary 
gauge theory with background field strength $B_{ij}$. 
In fact, the unitary operator $U^{(r)}$ cannot be uniquely 
determined. In the papers \refs{\hiko,\Sugino}, a unitary operator $e^{M^{(r)}}$ 
was proposed to be $M^{(r)}=(i / 2)\theta^{ij}{p_L}_i^{(r)}{p_R}_j^{(r)}$, 
where ${p_L}_i$ is given by the integration of the momentum $P_i(\sigma)$ over 
half a string $(0, \pi/2)$ and ${p_R}_i$ over $(\pi/2, \pi)$. Upon the unitary 
transformation, we obtain the transformed BRS charge $e^{M}\QB e^{-M}$ in the 
kinetic term and find that it is divergent. 
It seems that the divergence comes from the midpoint of open strings: 
$\sigma=\pi/2$. Since the transformed BRS charge should not depend 
on the types of the string interactions, it is natural to seek candidates for 
the operator $M$ which can be used in other string field theories, as well as 
in Witten's string field theory. As we argued in our paper \hiko, 
another operator $M$ has been proposed to be 
$M=-{i\o4}\int_0^\pi d\sigma\int_0^\pi d\sigma'
\epsilon(\sigma-\sigma')\theta^{ij}P_i(\sigma)P_j(\sigma')
$, which we called $\tilde{M}$ in \hiko. 
By making use of this operator $M$, the transformed BRS charge $e^{M}\QB e^{-M}$ 
can be found to be finite, as we will see below. 

Thus, if we perform the unitary transformation $\Psi\rightarrow{e^{-M}}\Psi$, 
we may expect to obtain another string field theory in the same background 
$B$-field, which gives an ordinary gauge theory with the background field 
strength $B_{ij}$ in the infrared. The aim of this paper is to show that this 
is indeed the case. To this end, it is appropriate to use an open-closed 
string field theory, because we have closed string fields as dynamical 
variables in such a theory. Since the antisymmetric tensor $B_{ij}$ is a 
component field of closed string fields, it is easy to understand the background 
$B$-field as the condensation of closed string fields. Thus, we can obtain 
an open-closed string field theory in the background $B$-field from the 
theory in the absence of the $B$-field by shifting closed string 
fields by the vacuum expectation value of the dynamical $B$-field contained in 
closed string fields. In this paper, we will use an open-closed string field 
theory recently given by Asakawa, Kugo, and one of the authors 
\refs{\KT\AKTa-\AKTb}.

The operator $M=-{i\o4}\int_0^\pi d\sigma\int_0^\pi d\sigma'
\epsilon(\sigma-\sigma')\theta^{ij}P_i(\sigma)P_j(\sigma')$ can be applied to 
the open-closed string field theory of the papers \refs{\KT\AKTa-\AKTb}. 
Indeed, we can express all the interactions of the string field theory 
in the $B$-field by using this operator $M$ and the vertices of the string 
field theory without the $B$-field, in the same way as we have done in Witten's 
string field theory \refs{\hiko,\Sugino}. In this case, we also need 
the interactions between open string fields and closed string fields. 
Therefore, one of the new results in our construction is 
the interaction between open strings and closed strings in the presence of the 
background $B$-field. In this paper, one of our main results is to show 
that this theory can be transformed into the theory obtained by the condensation 
of closed string fields by the above transformation 
$\Psi\rightarrow{e^{-M}}\Psi$. 

This paper is organized as follows: 
In section $2$, we construct the open-closed string field theory in the 
$B$-field by solving the overlapping conditions to obtain the string vertices. 
In section $3$, we obtain another string field theory in the same background as 
the theory in section $2$ by utilizing the condensation of closed string 
fields. In section $4$, we discuss the transformation of 
the string field theory in section $2$ into the theory in section $3$ 
by using the unitary transformation. Section $5$ is devoted to discussion. 
In the appendix, we solve the overlapping conditions for the open-closed string 
transition vertex and the open-open-closed string vertex 
in the background $B$-field to obtain these vertices.


\newsec{Open-Closed String Field Theory in the Background $B$-Field}

Some covariant string field theories have been constructed as covariantized 
light-cone gauge string field theories in \refs{\HIKKO\HIKKOclosed-\NeveuWest}. 
The theories include unphysical parameters called string length parameters 
$\alpha$, which are necessary to describe the joining-splitting vertices 
instead of the light-cone momentum $p^{+}$ in the light-cone gauge string field 
theory \refs{\KakuKikkawa,\Mandelstam}. 
In order to avoid such unphysical parameters, 
though we have to give up a manifestly Lorentz covariant formulation, 
we can use the so-called $\alpha=p^{+}$ HIKKO theories \KugoZwie, 
where we eliminate $p^{+}$ and identify $\alpha$ as $p^{+}$. 
This formulation still retains gauge invariance and is distinct from 
the light-cone gauge string field theory in this respect. 
In the series of papers \refs{\KT\AKTa-\AKTb}, an open-closed string 
field theory has been constructed in the $\alpha=p^{+}$ HIKKO formulation.

In this paper, we will study this theory in a constant background $B$-field. 
In general, the gauge invariance of this system is subtle due to the 
divergence from the dilaton tadpoles. 
Although, as shown in \refs{\AKTa,\AKTb}, gauge invariance can be 
maintained by the cancellation of the dilaton tadpoles from the disk amplitude 
and the crosscap amplitude, we will only consider the oriented sector 
in the theory. Since D-branes and orientifold plans are the sources of 
the disk and the crosscap, respectively, we will restrict 
our study to D-branes, despite the importance of orientifold planes 
in maintaining the full gauge invariance of the theory
\foot{We could consider another approach to a gauge 
invariant open-closed string
field theory \KugoSuppl, which is constructed in a curved background with 
gauge invariance ensured by the Fishler-Susskind mechanism 
\FS. However, such a theory has not yet been fully constructed.}.

As we will show in this section, the vertices of our string field theory 
in the $B$-field can be obtained by replacing open string 
fields ${\ket{\A}}$ with $e^{M}\hat{\ket{\A}}$ in the string field 
theory with no $B$-field, where $M$ is a particular operator. 
The aim of this section is to explain the derivation of this prescription.

Before we construct the string field theory in our background, 
let us recall some facts about the first-quantized open string theory 
in this background. The worldsheet action of open strings is 
\eqn\WSaction{
S=-{1\over4\pi\alpha'} \int d\tau\,\int^{\pi}_{0} d\sigma
\left(g_{ij}\eta^{ab}\partial_a \hX^i \partial_b \hX^j
-2\pi\alpha'B_{ij}\epsilon^{ab}
\partial_a \hX^i \partial_b \hX^j \right),
}
where $g_{ij}$ is a constant background metric and $B_{ij}$ is the constant 
background antisymmetric field. If the Dirichlet boundary 
condition is not chosen for all the directions of the string coordinates, 
the boundary condition 
can be seen to be 
$g_{ij}\d_{\sigma}{\hX^j}+ (2\pi\alpha') B_{ij}\d_{\tau}{\hX}^j =0$ 
at $\sigma=0,\,\pi$.
For simplicity, we will impose this boundary condition 
on all the string coordinates $\hX^i(\tau,\sigma)$.
The conjugate momenta $P_i(\sigma)$ are given by
$(1/2\pi\alpha')g_{ij}\d_{\tau}\hX^j(\sigma)
+B_{ij}\d_{\sigma}\hX^j(\sigma)$.

Applying the Dirac quantization method to this system 
\refs{\ChuHo,\Jabbari} (see also the appendix in \hiko), we obtain
the commutation relations
\eqn\CCR{\eqalign{
\left[\hX^i(\sigma),\ P_j(\sigma')\right]&=i\,
\delta^i_j\,{\delta}_c(\sigma,\sigma'), 
\qquad\left[P_i(\sigma),\ P_j(\sigma')\right]=0, 
\cr
&\left[\hX^i(\sigma),\ \hX^j(\sigma')\right]=
i\theta^{ij}\left(\delta_{\sigma,0}\delta_{\sigma',0}
-\delta_{\sigma,\pi}\delta_{\sigma',\pi}\right),
\cr
}}
where the delta function ${\delta}_c(\sigma,\sigma')$ is 
$(1/\pi)\sum^{\infty}_{n=-\infty}\cos(n\sigma)\cos(n\sigma')$
and $\delta_{\sigma,\sigma'}$ is the Kronecker delta.
The noncommutative parameter $\theta^{ij}$ is given by 
$[(g+2\pi\alpha'B)^{-1}\,B\,(g-2\pi\alpha'B)^{-1}]^{ij}$.
When turning off the $B$-field, we have the usual oscillator expansion 
of the string coordinates and the momenta
\eqn\openXP{
X^i(\sigma)=x^i+l_s\sum_{n\neq 0}{i\o n}
\alpha_n^i\cos(n\sigma),
\qquad
P_i(\sigma)={1\o \pi l_s}\sum G_{ij}\alpha^j_n\cos(n\sigma),
}
with $l_s=\sqrt{2\alpha'}$. 
In the presence of the $B$-field, the string coordinates become
$\hX^i(\sigma)=X^i(\sigma)+\theta^{ij} Q_j(\sigma)$,
where the open string metric $G_{ij}$ is 
$[g-(2\pi\alpha')^2Bg^{-1}B]_{ij}$,
while the momenta remain unchanged. The change $Q^i(\sigma)$ is given by 
$$
Q_i(\sigma)={1\o \pi}p_i\left(\sigma-{\pi \o 2}\right)
+{1\o \pi l_s}\sum_{n\neq 0}{1\o n}G_{ij}\alpha^j_n\sin(n\sigma)
$$
and is related to $P_i(\sigma)$ by 
\eqn\QP{
Q_i(\sigma)={1\o 2}\int^\pi_0 d\sigma' \epsilon(\sigma-\sigma')P_i(\sigma').
}
From \CCR, we can obtain the commutation relations of these oscillators 
\eqn\ModeCR{
[x^i,\,p_j]=i\delta^i_j,\ \ \ \left[\alpha_m^i,\,
\alpha^j_n\right]=m\delta_{m+n}G^{ij}.
}

As we mentioned at the beginning of this section, 
in the resulting string field theory with the $B$-field,
we will find that open string fields ${\ket{\A}}$ of the theory 
without the $B$-field are multiplied by the factor $e^{M}$. 
Therefore, it is appropriate to define this operator $M$ 
before we give the vertices of the string field theory. 
The operator $M$ is given by 
\eqn\M{
M=-{i\o 4}\int^\pi_0d\sigma\int^\pi_0d\sigma'
\epsilon(\sigma-\sigma')\theta^{ij}P_i(\sigma)P_j(\sigma'),
}
where $\epsilon(\sigma)$ is a step function which is $1$ for $\sigma>0$
and $-1$ for $\sigma<0$. When solving the overlapping conditions to obtain 
the vertices, we will find that the usual vertices in \refs{\KT\AKTa-\AKTb} 
fail to satisfy those conditions in our background. 
But if we put the operator $e^{M}$ on each leg of the open strings, 
the resulting vertices satisfy them. This is similar to what we have done 
for the midpoint interaction of Witten's theory in \refs{\hiko,\Sugino}
\foot{In Witten's theory, we can use another operator to make the vertices 
satisfy the overlapping conditions. See \refs{\hiko,\Sugino} 
for further details.
}.

For closed strings, since we are considering a flat Minkowski spacetime, 
the $B$-field cannot alter the dynamics of this system. Indeed, though we have 
the same worldsheet action as \WSaction, except that $\sigma$ runs from 
$0$ to $2\pi$, not to $\pi$, the term including the $B$-field can be made into 
a surface term and can thus be discarded. But later it will be convenient to 
have kept this term in the action. 
In the language of the Hamiltonian formulation, 
keeping the surface term means that we have alternative canonical variables 
$X^i(\sigma)$ and $P_i(\sigma)$; from the action, the momenta $P_i(\sigma)$ 
can be seen to be $(1/2\pi\alpha')g_{ij}\dot{X}^i(\sigma)+
B_{ij}X^{'j}(\sigma)$.

Since the string coordinates obey the usual equation of motion and 
the periodic boundary condition is imposed on them, 
they have the usual oscillator expansion
\eqn\XcMode{
X^i(\sigma)=x^i+{l_s\o 2}\sum_{n\neq 0}{i\o n}
\left(\alpha_n^i e^{-i n\sigma}+\bar{\alpha}_n^i e^{i n\sigma}
\right).
}
It is obvious that the combination 
$(P_i-B_{ij}X^{'j})$ 
has the ordinary expansion of the usual momenta, therefore we obtain
\eqn\XPcMode{
P_i(\sigma)={1\o2\pi}p_i
+{1\o 2\pi l_s}\sum_{n\not=0}
\left[\left(g_{ij}+b_{ij}\right)\alpha_n^j e^{-i n\sigma}
+\left(g_{ij}-b_{ij}\right)\bar{\alpha}_n^j e^{i n\sigma}\right],
}
where $b_{ij}=(2\pi\alpha')B_{ij}$.
By using the usual commutation relations of the string coordinates and 
the momenta, we can verify that the oscillators satisfy
\eqn\CModeCR{
[x^i,\,p_j]=i\delta^i_j,\ \ \ \left[\alpha_m^i,\,
\alpha^j_n\right]=m\delta_{m+n}g^{ij},\ \ \ 
\left[\bar{\alpha}_m^i,\,
\bar{\alpha}^j_n\right]=m\delta_{m+n}g^{ij}.
}
We will also find it useful to define for closed strings 
the counterpart of the operator $M$ 
\eqn\Mclose{
M=-{i\o 4}\int_0^{2\pi}d\sigma \int_0^{2\pi}d\sigma'
\epsilon(\sigma-\sigma')\theta^{ij}P_i(\sigma)P_j(\sigma').
}

The BRS charges in this background are needed to obtain the kinetic terms of 
our string field theory. 
In the string field theory, we have the kinetic term of open string 
fields $\hat{\ket{\Psi}}$ and that of closed string fields $\ket{\Phi}$. 
Because we have the term including the $B$-field in the action for closed 
strings, the BRS charge for closed strings in our background is the same form 
as that for open strings, when we write them 
in terms of the string coordinates and the momenta. 
The ghost part of the BRS charges is not different from 
the usual one. The matter part can be given by using the energy-momentum tensor 
\eqn\hTXP{\eqalign{
\hat{T}_{\pm\pm}
&={1\o 4\alpha'}g^{ij}
\left[(2\pi\alpha')P_i\pm\left(g_{ik}\mp b_{ik}\right)
\hat{X}^{\prime{k}}\right]
\left[(2\pi\alpha')P_j\pm\left(g_{jl}\mp b_{jl}\right)
\hat{X}^{\prime{l}}\right]
\cr
&={1\o 4\alpha'}G_{ij}
\left[\hat{X}^{\prime{i}}\pm(2\pi\alpha')\left(G^{ik}\mp
{\theta^{ik}\o2\pi\alpha'}\right)P_k\right]
\left[\hat{X}^{\prime{j}}\pm(2\pi\alpha')\left(G^{jl}\mp
{\theta^{jl}\o2\pi\alpha'}\right)P_l\right],
\cr}
}
where the prime denotes the differentiation with respect to $\sigma$. 
Here, the above expression \hTXP\ can be applied to the energy-momentum tensor 
only for open strings. In order to obtain the energy-momentum tensor 
for closed strings, we have to replace the string coordinates $\hX^i(\sigma)$ 
with $X^i(\sigma)$ in \XcMode\ and take away the hat on the energy-momentum 
tensor. 
Noting the combination $(2\pi\alpha')g^{ij}(P_i-B_{ij}X^{'j})$ in 
the former expression of the energy-momentum tensor in \hTXP, 
we can see that ${T}_{\pm\pm}$ for closed strings has the usual oscillator 
expansion. We can use the latter expression of $\hat{T}_{\pm\pm}$ 
for open strings to understand that $\hat{T}_{\pm\pm}$ 
also has the usual oscillator expression 
except that the closed string metric $g_{ij}$ is replaced by the open string 
metric $G_{ij}$, as we mentioned in the introduction.

Now, let us consider our open-closed string field theory in the background 
$B$-field. In the absence of the $B$-field, the action $S$ of 
the oriented open-closed string field theory is

\eqn\action{\eqalign{
-&{1\o 2}{\bra{\Psi}}\QB{\ket{\Psi}}
-{1\o 2}\bra{\Phi}\QB(b_0^-{\cal P})\ket{\Phi}
+{g\o 3}\Vo{1,2,3} {\ket{\Psi}}_{321}
+{g^2\o 4}\Vfo{1,2,3,4}{\ket{\Psi}}_{4321}
\cr
&+{g^2\o 3!}\Vc{1^\rc,2^\rc,3^\rc}\ket{\Phi}_{321}
+g \U{1,2^\rc}\ket{\Phi}_2{\ket{\Psi}}_1
+{g^2 \o 2}\Uomg{1,2,3^\rc}\ket{\Phi}_3{\ket{\Psi}}_{21},
}}
as has been given in \refs{\AKTa}
\foot{In this action, the coefficients $x_\rc$, $x_u$ and $x_\Omega$ given 
in \refs{\AKTa,\AKTb} are included in their respective vertices. 
Therefore, the definition of the vertices is different from that in 
\refs{\AKTa,\AKTb} by these multiplicative factors.}.
Here $g$ is the open string coupling constant. 
The sign $\hat{\ket{\Psi}}$ will be used for an open string field 
in our theory with $B$-field.
The multiple products $\ket{\Phi}_n\cdots\ket{\Phi}_2\ket{\Phi}_1$ 
of string fields are denoted by $\ket{\Phi}_{n\cdots21}$.
In \action, we omitted the integrations over the zero modes 
of the momenta. In the $\alpha=p^{+}$ HIKKO formulation, we identify 
the $\alpha$ parameter as $p^{+}$ for closed strings and as $2p^{+}$ 
for open strings.

We have five interaction vertices in this theory. 
These vertices can be obtained up to an overall normalization by 
solving the overlap condition. 
Similarly, in our background the string vertices 
are given by the overlapping condition.  
For example, the open three-string vertex $\hVo{1,2,3}$
in the case where $\alpha_1,\,\alpha_2>0,\ \ \alpha_3<0$ 
is explicitly given by the overlapping condition
\eqn\hVocon{\eqalign{
&\hVo{3,2,1}\left\{\hat{\phi}^{(1)}\left({\sigma_1}\right)
-\hat{\phi}^{(3)}\left({\sigma_3}\right)
\right\}=0,\ \ \ (0<\sigma<\pi \alpha_1),
\cr
&\hVo{3,2,1}\left\{\hat{\phi}^{(2)}\left({\sigma_2}\right)
-\hat{\phi}^{(3)}\left({\sigma_3}\right)
\right\}=0,\ \ \ (\pi \alpha_1< \sigma < \pi |\alpha_3|),
}}
where $\hat{\phi_r}^{(r)}(\sigma_r)$ denotes $\hX^{i\,(r)}(\sigma_r)$
or $(1/\alpha_r)P_i^{(r)}(\sigma_r)$ of the $r$-th open string.
Here we use the following parameters: $\sigma_1=\sigma/\alpha_1$; 
$\sigma_2=(\sigma-\pi \alpha_1)/\alpha_2$; 
$\sigma_3=(\pi|\alpha_3|-\sigma)/|\alpha_3|$. 
Likewise, the usual open three-string vertex $\Vo{1,2,3}$ is specified by
\eqn\Vocon{\eqalign{
&\Vo{3,2,1}\left\{\phi^{(1)}\left({\sigma_1}\right)
-\phi^{(3)}\left({\sigma_3}\right)
\right\}=0,\ \ \ (0<\sigma<\pi \alpha_1),
\cr
&\Vo{3,2,1}\left\{\phi^{(2)}\left({\sigma_2}\right)
-\phi^{(3)}\left({\sigma_3}\right)
\right\}=0,\ \ \ (\pi \alpha_1< \sigma < \pi |\alpha_3|),
}}
where $\phi^{(r)}(\sigma_r)$ is the same as $\hat\phi^{(r)}(\sigma_r)$, 
but it denotes $X^{i\,(r)}(\sigma_r)$, not $\hX^{i\,(r)}(\sigma_r)$. 
This vertex $\Vo{1,2,3}$ can be used to solve the overlapping condition 
\hVocon\ to get our vertex $\hVo{1,2,3}$.
Replacing $X^{i\,(r)}(\sigma_r)$ with $\hX^{i\,(r)}(\sigma_r)$ in the left-hand 
side of \Vocon, we find that those string coordinates do not 
connect on the vertex by the difference $Q^{i\,(r)}(\sigma_r)$ between 
$\hX^{i\,(r)}(\sigma_r)$ and $X^{i\,(r)}(\sigma_r)$. Indeed, using \Vocon\ and 
\QP, 
we find 
\eqn\VoQcon{\eqalign{
&\Vo{1,2,3}\left\{
Q^{i\,(1)}\left({\sigma_1}\right)-
Q^{i\,(3)}\left({\sigma_3}\right)
\right\}
=\Vo{1,2,3}{1\o 2}G^{ij}p_j^{(2)},\ 
(0<\sigma<\pi \alpha_1)
\cr
&\Vo{1,2,3}\left\{
Q^{i\,(2)}\left({\sigma_2}\right)-
Q^{i\,(3)}\left({\sigma_3}\right)
\right\}
=\Vo{1,2,3}\left(-{1\o 2}\right)G^{ij}p_j^{(1)}.\ 
(\pi \alpha_1<\sigma<\pi|\alpha_3|)
}}
This equation \VoQcon\ helps us to find our open three-string vertex 
\eqn\VtVoP{
\hVo{1,2,3}=\Vo{1,2,3}\exp\left(-{i\o 2}\sum_{r<s}
\theta^{ij}p_i^{(r)}p_j^{(s)}\right),
}
which is the same expression as that in Witten's open string field theory, 
as has been found in \refs{\Sugino,\hiko}. 

The noncommutative factor $\exp[-(i/2)\sum_{r<s}\theta^{ij}p_i^{(r)}p_j^{(s)}]$ 
in \VtVoP\ can be rewritten with the operator $M$, as has been discussed 
in \hiko. In fact, changing the variables $\sigma$ in \M\ 
to the above $\sigma_3$, we find that 
the operator $M$ for the third string turns into 
\eqn\Mth{\eqalign{
M^{(3)} =&-{i\o 2}\int^{\pi\alpha_1}_0 d\sigma
\int^{\pi|\alpha_3|}_{\pi\alpha_1}d\sigma'
\theta^{ij}{1\o\alpha_3}P_i^{(3)}(\sigma_3)
{1\o\alpha_3}P_j^{(3)}(\sigma_3')
\cr
&+{i\o 4}\int^{\pi\alpha_1}_0d\sigma
\int^{\pi\alpha_1}_0d\sigma' \epsilon(\sigma-\sigma')\theta^{ij}
{1\o\alpha_3}P_i^{(3)}(\sigma_3)
{1\o\alpha_3}P_j^{(3)}(\sigma_3')
\cr
&+{i\o 4}\int^{\pi|\alpha_3|}_{\pi\alpha_1}d\sigma
\int^{\pi|\alpha_3|}_{\pi\alpha_1}d\sigma'
\epsilon(\sigma-\sigma')\theta^{ij}
{1\o\alpha_3}P_i^{(3)}(\sigma_3)
{1\o\alpha_3}P_j^{(3)}(\sigma_3').
}}
Putting this operator $M^{(3)}$ on the vertex $\hVo{1,2,3}$, 
due to the overlapping condition \hVocon, we can see that 
the second term in the right-hand side of \Mth\ can be converted to 
$M^{(1)}$ and the third term to $M^{(2)}$. Furthermore, the first term 
becomes $-(i/2)\theta^{ij}p_i^{(1)}p_j^{(2)}$, which can be made into
$-(i/2)\sum_{r<s}\theta^{ij}p_i^{(r)}p_j^{(s)}$ 
by using the momentum conservation. Thus, 
\eqn\VtVoM{
\hVo{1,2,3}=\Vo{1,2,3}\prod_{r=1}^3 e^{M^{(r)}}.
}
In the same manner, the open four-string vertex $\hVfo{1,2,3,4}$ can be 
found to be $\Vfo{1,2,3,4}\exp[-(1/2)\sum_{r<s}\theta^{ij}p_i^{(r)}p_j^{(s)}]$ 
and be rewritten into 
$$
\hVfo{1,2,3,4}=\Vfo{1,2,3,4}\prod_{r=1}^4 e^{M^{(r)}}.
$$
Also, we can easily verify that the closed three-string vertex remains intact: 
$$
\hVc{3^\rc,2^\rc,1^\rc}=\Vc{3^\rc,2^\rc,1^\rc}.
$$

The vertex $\U{1,2^\rc}$ gives the transition between an open string and a 
closed one, and $\Uomg{1,2,3^\rc}$ gives the interaction between two open 
strings and a closed one. Since the coordinates of 
the ends of open strings are noncommutative on D-branes and the $B$-field 
can only change the dynamics of closed strings through interaction with 
open strings, the modification of 
these vertices by the $B$-field are not given by the noncommutative factor 
$\exp[-(i/2)\sum_{r<s}\theta^{ij}p_i^{(r)}p_j^{(s)}]$. However, as 
discussed in the appendix, by taking advantage of the operator $M$ to solve 
the overlap conditions, we can obtain those vertices 
$$
\eqalign{
&\hU{2,1^\rc}=\U{2,1^\rc} e^{M^{(2)}},
\cr
&\hUomg{3,2,1^\rc}=\Uomg{3,2,1^\rc}e^{M^{(2)}}e^{M^{(3)}}.
}
$$

Thus, summing up our results about the vertices in our string field theory, 
we only have to substitute 
$e^{M}\hat{\ket{\Psi}}$ for $\ket{\Psi}$ in the vertices in the absence of 
the $B$-field to obtain all the vertices in our background. 
Therefore, the action of our string field theory is given by 
\eqn\haction{\eqalign{
\hat{S}=-&{1\o 2}\hat{\bra{\Psi}}\hQB\hat{\ket{\Psi}}
-{1\o 2}\bra{\Phi}\hQB(b_0^-{\cal P})\ket{\Phi}
\cr
&+{g\o 3}\hVo{1,2,3} \hat{\ket{\Psi}}_{321}
+{g^2\o 4}\hVfo{1,2,3,4}\hat{\ket{\Psi}}_{4321}
+{g^2\o 3!}\hVc{1^\rc,2^\rc,3^\rc}\ket{\Phi}_{321}
\cr
&+g \hU{1,2^\rc}\ket{\Phi}_2\hat{\ket{\Psi}}_1
+{g^2 \o 2}\hUomg{1,2,3^\rc}\ket{\Phi}_3\hat{\ket{\Psi}}_{21}.
}}
In the next section, we will obtain an alternative string field theory 
in the same background through the condensation of 
the $B$-field included in closed string fields $\ket{\Phi}$. 


\newsec{The Condensation of the $B$-Field}

In the previous section, we obtained the string field theory in the 
background $B$-field by constructing the vertices as the solutions of 
the overlapping conditions. 
The dynamical variables of this field theory are closed string fields as well 
as open string ones. Since the $B$-field is a component field of closed string 
fields, the background $B$-field can be obtained through the condensation of 
closed string fields. Therefore, beginning with the open-closed string field 
theory \action\ in the absence of the background $B$-field, and shifting closed 
string fields by the vacuum expectation value of the $B$-field, 
we obtain a string field theory in the same background as that 
in the previous section. The resulting theory is different from the theory 
in the previous section, because only the BRS charge is affected by the 
condensation. Unfortunately, it is difficult to shift closed string fields 
by a finite vacuum expectation value. Therefore, we will 
demonstrate the shifting by an infinitesimal amount in the string field theory. 
Then, integrating the infinitesimal change of the BRS charge under the shifting, 
we can find the string field theory in the finite $B$-field.

\subsec{The Background Independence of the String Vertices}

Before calculating the contraction of the string vertices with the shifted 
closed string fields, we need to understand what remains unchanged under the 
change of the background $B$-field: $B_{ij}\rightarrow B_{ij}+\dB_{ij}$, 
as Kugo and Zwiebach have done in \KugoZwie. Although they dealt with the closed 
string field theory compactified on a torus, 
we will follow their method and extend the assumption to our system of 
open strings as well as closed strings. Indeed, we assume that 
the string coordinates $X^i$ and their momenta $P_i$ of open strings and closed 
strings are background independent. In the previous section, we have dealt with 
the string coordinates $\hat{X}^i(\sigma)$. These coordinates satisfy 
the boundary condition: $\d_{\sigma}{\hX^j}-\theta^{ij}P_j =0$ 
at $\sigma=0,\,\pi$. If we assumed that $\hat{X}^i(\sigma)$ was background 
independent, it would be inconsistent with this boundary condition. 
To avoid this inconsistency, since 
$\d_{\sigma}\hX^i=\d_{\sigma}X^i+\theta^{ij}P_j$, we should assume that, 
rather than $\hX^i(\sigma)$, the string coordinates $X^i(\sigma)$ of open 
strings are background independent, as is the momenta $P_j(\sigma)$.

By applying this assumption to the energy-momentum tensor of closed strings, 
as we can see from \hTXP, under the infinitesimal change of the 
background $B$-field, we can obtain the variation of the energy-momentum tensor
\eqn\delEM{
\delta T_{\pm\pm}(\sigma)=-{2\pi\a'\o2\a'}g^{ik}\delta{B}_{kj}
\left(P_i(\sigma)-B_{il}{X'}^l(\sigma)\right){X'}^j(\sigma).
}
Since the BRS charge 
for closed strings is given by 
\eqn\cBRS{
\QB=\int^{2\pi}_0{d\sigma\o2\pi}\left[c(\sigma)T_{++}(\sigma)+
\bar{c}(\sigma)T_{--}(\sigma)\right], 
}
its variation turns out to be
\eqn\delQB{
\delta\QB=-\int^{2\pi}_0d\sigma\hf\left(c(\sigma)+\bar{c}(\sigma)\right) 
g^{ik}\delta{B}_{kj}\left(P_i(\sigma)-B_{il}{X'}^l(\sigma)\right){X'}^j(\sigma).
}
If we had dealt with the BRS charge for open strings in the above and 
had replaced the string coordinates $\hX^i(\sigma)$ with the 
background-independent coordinates $X^i(\sigma)$, we would have obtained the 
variation of the BRS charge in the same form as \delQB\ except for the 
integration over $(0,\pi)$ instead of $(0,2\pi)$. 

By using \XcMode\ and \XPcMode, in terms of the string coordinates and 
the momenta, we can express the oscillators for closed strings as 
\eqn\oscillator{\eqalign{
&\a^i_n(B)=\int^{2\pi}_0{d\sigma\o2\pi\l}e^{in\sigma}g^{ij}
\left[(2\pi\a')P_j-(2\pi\a')B_{jk}{X'}^k+g_{jk}{X'}^k\right],
\cr
&\bar\a^i_n(B)=\int^{2\pi}_0{d\sigma\o2\pi\l}e^{-in\sigma}g^{ij}
\left[(2\pi\a')P_j-(2\pi\a')B_{jk}{X'}^k-g_{jk}{X'}^k\right].
}}
Here we labeled the oscillators with the letter $B$ as 
$(\a^i_n(B), \bar\a^i_n(B))$ 
to explicitly show their dependence on the background $B$-field. 
These oscillators $(\a^i_n(B), \bar\a^i_n(B))$ can be related to the 
oscillators $(\a^i_n(0), \bar\a^i_n(0))$ in the absence of the $B$-field 
by using the operator
\eqn\beta{
{\cal B}=-{i\o2}\int^{2\pi}_0d\sigma B_{ij}{X'}^i(\sigma)X^j(\sigma).
}
In fact, since 
$e^{{\cal B}}P_i(\sigma)e^{-{\cal B}}=P_i(\sigma)-B_{ik}{X'}^j(\sigma)$,
we can see from \oscillator\ that 
\eqn\trfbetaalpha{
e^{{\cal B}}\left(\alpha^i_n(B), \bar\alpha^i_n(B)\right)e^{-{\cal B}}
=\left(\alpha^i_n(0), \bar\alpha^i_n(0)\right). 
}
Also, the $SL(2,\IC)$ invariant Fock vacuum depends upon the background 
$B$-field. Since the Fock vacuum is defined by 
$\left(\alpha^i_n(B), \bar\alpha^i_n(B)\right)\ket{0;0}_{B}=0$
for $n\geq1$, it is easy to verify that 
\eqn\trfFock{
\ket{0;0}_{B}=e^{{\cal B}}\ket{0;0}_{0}.
}
Here, we also labeled the Fock vacua with the letter $B$ to show its dependence 
on the $B$-field. 

In the next subsection, we will try to find the string field theory 
in the background $B$-field by gradually increasing the amount of the background 
$B$-field. We will do this by taking advantage of the condensation of closed 
string fields. This condensation turns out to only affect the kinetic term, 
{\it i.e.}, the BRS charge, but not the interaction terms. Therefore, in this 
paper, we will assume that the string vertices are background independent. 
This is also supported by the fact that the string vertices are specified 
up to an overall normalization factor by the overlapping conditions which 
are written in terms of the string coordinates $X^i(\sigma)$ and their momenta 
$P_i(\sigma)$.

\subsec{The Calculation for the Shifting of Closed String Fields}
An infinitesimal vacuum expectation value of the $B$-field 
contained in closed string fields $\ket{\Phi}$ is given by 
\eqn\delPhi{
\ket{\delta\Phi}
={1\o x_cg^2}c_0^{-} \dB_{ij}\,\alpha_{-1}^i(B){\bar{\alpha}}_{-1}^j(B)
\ket{1,1}_B(2\pi)^{26} \delta^{26}(p), 
}
where $\dB_{ij}$ is the infinitesimal constant background of the $B$-field and 
$x_c$ is the numerical factor \AKTa\ included in the closed three-string vertex. 
In addition, $c^{-}_0$ is the zero mode of the reparametrization ghost 
and is given by the combination $c_0-\bar{c}_0$. $\ket{1,1}_B$ denotes the state 
$c_1\bar{c}_1\ket{0,0}_B$.
By shifting closed string fields as 
$\ket{\Phi}\rightarrow\ket{\Phi}+\ket{\delta\Phi}$ in \action, we can obtain 
an open-closed string field theory in the background $B$-field. To calculate 
this shifting in \action, we will follow the prescription \LPP\ 
given by LeClair, Peskin, and Preitschopf, 
which will facilitate our calculation. 

The essential point of the prescription \LPP\ is to take advantage of 
conformal field theory. The LPP vertices give the relation between vertices in 
string field theory and correlation functions in the corresponding conformal 
field theory, as we will see below. In the action \action, in terms of the 
LPP vertices $\bra{v(n,\cdots,2,1)}$, we can write the five vertices \AKTa\ as 
\eqn\defvertex{\eqalign{
&\Vo{1,2,3} = \vo{1,2,3},
\cr
&\Vc{1^\rc,2^\rc,3^\rc} = x_c\vc{1^\rc,2^\rc,3^\rc}\prod_{r=1^\rc,2^\rc,3^\rc}
(b_0^- {\cal P})^{(r)},
\cr
&\U{1,2^\rc} = x_u\u{1,2^\rc}(b_0^-{\cal P})^{(2^\rc)},
\cr
&\Vfo{1,2,3,4} = x_4\int_{\sigma_i}^{\sigma_f}d\sigma_0^{(1)}
\vfo{1,2,3,4;\sigma_0^{(1)}}b_{\sigma_0^{(1)}},
\cr
&\Uomg{1,2,3^\rc} = x_{\Omega}\int_{\sigma_i}^{\sigma_f}d\sigma_0^{(1)}
\uomg{1,2,3^\rc;\sigma_0^{(1)}}b_{\sigma_0^{(1)}}
(b_0^-{\cal P})^{(3^\rc)}, 
}}
where the vertices denoted by the small letters are the LPP vertices. 
The parameter $\sigma_0$ is a moduli parameter of the vertices and ${\cal P}$ is 
the operator which projects onto the $L_0=\bar{L}_0$ sector for closed strings.
The necessity of the inserted anti-ghosts operator $b$ is explained 
in \refs{\AGGMV,\AKTa}.
In particular, $b^{-}_{0}$ is given by $(b_{0}-\bar{b}_{0})/2$. 

These LPP vertices are defined by
\eqn\LPPvertex{
\bra{v(n,\cdots,2,1)}\ket{A_1}_1\ket{A_2}_2\cdots\ket{A_n}_n =
\left\langle h_1[{\cal O}_{A_1}]h_2[{\cal O}_{A_2}]\cdots h_n[{\cal O}_{A_n}]
\right\rangle
}
where the right-hand side denotes the correlation function of the operators 
${\cal O}_{A_r}$ transformed by the conformal mappings $h_r$ in the conformal 
field theory.  
Here, the string fields $\ket{A_r}_r$ are ${\cal O}_A \ket{0}_r$,
and each string field $\ket{A_r}_r$ corresponds to a unit disc 
$|w_r|\leq 1$ of which the corresponding operator ${\cal O}_{A_r}$ is 
inserted at the origin. These unit discs are mapped by $h_r$ into a single 
complex $z$-plane to glue the strings. 
In particular, if the operator ${\cal O}$ 
is a primary field $\phi(w,\bar{w})$ of weight $(d,\bar{d})$,
it is transformed by the conformal mapping $z=h(w)$ into 
\eqn\hO{
h[\phi(w,\bar{w})]=\left({dh(w)\o dw}\right)^d
\left({d\bar{h}(\bar{w})\o d\bar{w}}\right)^{\bar{d}}
\phi\left(h(w),\bar{h}(\bar{w})\right).
}
As a simple example, the reflector for open strings is defined by
\eqn\orefI{
\bra{R^\ro(2,1)}\ket{A}_1\ket{B}_2
=\left<I[{\cal O}_A] {\cal O}_B \right>,
}
where $I(z)=-1/z$ \LPP. For our calculation later, it will be useful to 
introduce another definition of the reflector 
\eqn\oref{
\bra{R^\ro(2,1)}\ket{A}_1\ket{B}_2
=\left<h_1^R[{\cal O}_A] h_2^R[{\cal O}_B] \right>.
}
These conformal mappings $h_r^R(w_r)$ are given by the Mandelstam mapping
$\rho = \alpha_1 \ln(z-Z_1) + \alpha_2 \ln(z-Z_2)$
where $\alpha_1+\alpha_2=0$. 
Here we use an intermediate $\rho$-plane: $w_r=e^{\rho_r}$;
$\rho=\alpha_1\rho_1+\rho_0$ for the first string and  
$\rho=\alpha_2 \rho_2 +\rho_0+i\pi\alpha_1$ for the second string. 
$\rho_0$ is the interaction point. 

Thus, in order to make use of this technique, we have to go to the Euclidean 
signature on the worldsheet. In the conformal field theory language, 
the string coordinates of closed strings are given by 
$\partial{X}^i(w)=(l_s/2i)\sum^{\infty}_{n=-\infty}
\alpha^i_nw^{-n-1}$ and $\bar{\partial}{X}^i(\bar{w})=(l_s/2i)
\sum^{\infty}_{n=-\infty}\bar{\alpha}^i_n\bar{w}^{-n-1}$. 
For the reparametrization ghost, $c(w)=\sum^{\infty}_{n=-\infty}
c_nw^{-n+1}$ and $\bar{c}(\bar{w})=\sum^{\infty}_{n=-\infty}
\bar{c}_n\bar{w}^{-n+1}$. In terms of these variables, 
$\ket{\delta\Phi}$ can be written as
\eqn\CFTdelPhi{
\ket{\delta\Phi}
=-c_0^{-}\left({2\o\a'x_cg^2}\right)\dB_{ij}\,\lim_{w\rightarrow 0}
c\,\bar{c}\,\partial{X}^i\bar{\partial}{X}^j(w,\bar{w})
\ket{0,0}_B(2\pi)^{26}\delta^{26}(p). 
}
To accurately formulate our calculation, it is useful to introduce 
a `regularized' state
\eqn\regPhi{
\ket{B_{\epsilon}^{ij}}_B=\lim_{w\rightarrow 0} 
c\,\bar{c}\,\partial{X}^i\bar{\partial}{X}^j(w,\bar{w})
\ket{0,0}_B(2\pi)^{26}\delta_{\epsilon}^{26}(p), 
}
where $\delta^{26}_\epsilon(p)=(1/2)
[\delta(p^+-\epsilon)+\delta(p^++\epsilon)]\delta^{25}(p)$. 
Therefore, we can see that 
$c_0^{-}\dB_{ij}\ket{B_{\epsilon}^{ij}}
\rightarrow-(\a'x_cg^2/2)\ket{\delta\Phi}$ in the limit $\epsilon\rightarrow0$.

Now, let us calculate the closed three-string vertex $\Vc{1^\rc,2^\rc,3^\rc}$
inserted by the shift $\ket{\delta\Phi}$ by following the LPP prescription. 
The conformal mappings $h_r$ for the closed three-string vertex
are given by the Mandelstam mapping \refs{\Mandelstam,\HIKKOclosed}
\eqn\Mandelclosed{
\rho=\sum^{3}_{r=1}
\alpha_r \ln(z-Z_r)
=\alpha_r \ln w_r +\rho_0^{(r)},
}
where string length parameters satisfy $\sum_{r=1}^3\alpha_r=0$.
For our calculation, we only need the vertex with a particular configuration of 
strings where $\alpha_2,\alpha_3 >0$ and $\alpha_1<0$. In the Madelstam mapping 
\Mandelclosed\ of the configuration, $\rho_0^{(1)}=\rho_0+i\pi\alpha_{3}$; 
$\rho_0^{(2)}=\rho_0-i\pi\alpha_{2}$; $\rho_0^{(3)}=\rho_0$, 
where $\rho_0$ is the interaction point given by $\rho_0=\rho(z_0)$; 
$z_0$ is defined by $(\partial{\rho}/\partial{z})(z=z_0)=0$.
The variable $Z_r$ is the position of the operator of the $r$-th 
string on the complex $z$-plane and are specified by $Z_r=h_r(w_r=0)$.


Using the property of the LPP vertex, we find that
\eqn\vcAB{
\vc{1^\rc,2^\rc,3^\rc}\ket{B_{\epsilon}^{ij}}_3 \ket{A}_2 \ket{B}_1 
= \left\langle c\,\bar{c}\,\partial{X}^i\bar{\partial}{X}^j(Z_3,\bar{Z}_3)\,
h_2^{\epsilon}[{\cal O}_A]\,h_1^{\epsilon}[{\cal O}_B]\right\rangle,
}
where we introduce the label $\epsilon$ for the above mappings $h_r$ to 
remind us that the light-cone momentum $p^{+}$ of the third string is 
infinitesimal: $\alpha_3\sim\epsilon$. In the right-hand side of \vcAB, 
since the operator $c\bar{c}\partial{X}^i\bar{\partial}{X}^j$ 
contracted with the antisymmetric tensor $\delta{B}_{ij}$ 
in the shift is a primary field with its weight $(0,0)$ and $h_3(w_3=0)=Z_3$, 
we have used the fact that $\lim_{w\rightarrow 0}h_3[c\bar{c}
\delta{B}_{ij}\partial{X}^i\bar{\partial}{X}^j(w,\bar{w})]=c\bar{c}
\delta{B}_{ij}\partial{X}^i\bar{\partial}{X}^j(Z_3,\bar{Z}_3)$. 


In the limit $\alpha_3\rightarrow0$, we can see from \Mandelclosed\ that 
the mappings $h^{\epsilon}_r$ become $h^{R}_r$, 
which are the conformal mappings to give the reflector for closed strings: 
the counterpart of the mapping associated with 
the reflector \oref\ for open strings. 
The insertion point $Z_3$ goes to the interaction point $z_0$ on the $z$-plane. 
Therefore, the operator 
$c\bar{c}\partial{X}^i\bar{\partial}{X}^j(Z_3,\bar{Z}_3)$ in the right-hand 
side of \vcAB\ becomes 
$c\bar{c}\partial{X}^i\bar{\partial}{X}^j(z_0,\bar{z}_0)$. Since the 
interaction $z_0$ corresponds to the point $w_1=1$ of the first string, we 
can regard the operator 
$c\bar{c}\partial{X}^i\bar{\partial}{X}^j(z_0,\bar{z}_0)$ as 
$h^{R}_1[c\bar{c}\partial{X}^i\bar{\partial}{X}^j(w=1,\bar{w}=1)]$.
Thus, we find that 
\eqn\vcABlimit{
\vc{1^\rc,2^\rc,3^\rc}\ket{B_{\epsilon}^{ij}}_3 
=\bra{R^\rc(1^\rc,2^\rc)}\,
\left(c\bar{c}\partial{X}^i\bar{\partial}{X}^j\right)^{(1^\rc)}(w=1,\bar{w}=1),
}
where we dropped the states $\ket{A}_2$ and $\ket{B}_1$, 
when the both sides are contracted with $\delta{B}_{ij}$. 
Since we made no assumptions in choosing these states, \vcABlimit\ should hold 
identically. 

When we contract the closed three-string vertex $\Vc{1^\rc,2^\rc,3^\rc}$ 
with the shift $\ket{\delta\Phi}$, we can see that the zero mode $b^{-}_0$ 
from the vertex cancels $c^{-}_0$ from the shift by their commutation relation 
$\{b^{-}_0,c^{-}_0\}=1$ and that the product of 
$\vc{1^\rc,2^\rc,3^\rc}\ket{B_{\epsilon}^{ij}}_3$ and 
$({\cal P}b^{-}_0)^{(1)}({\cal P}b^{-}_0)^{(2)}$ are left. 
By \vcABlimit, the former factor becomes the reflector times the operator 
$-(2/\a'x_cg^2)\dB_{ij}c\bar{c}\partial{X}^i\bar{\partial}{X}^j$ of 
the first string. We make the projection ${\cal P}$ and 
the anti-ghost zero mode $b^{-}_0$ of the second string pass through 
the operator $c\bar{c}\partial{X}^i\bar{\partial}{X}^j$. 
After contracting the anti-ghost zero mode $b^{-}_0$ with the ghost operators 
$c(\sigma)\bar{c}(\sigma)$, 
we find that $\Vc{1^\rc,2^\rc,3^\rc}\ket{\delta\Phi}_3$ becomes 
\eqn\preVceps{
\bra{R^\rc(1^\rc,2^\rc)}
\left({1\o \a'g^2}\right){\cal P}^{(1^\rc)}(c^{(1)}+\bar{c}^{(1)})
\,\dB_{ij}\,\partial{X}^{i(1)}\,\bar{\partial}{X}^{j(1)}(w=1,\bar{w}=1)\,
(b_0^-{\cal P})^{(1^\rc)}. 
}
Since the projection operator ${\cal P}$ is given by 
$(1/2\pi)\int^{2\pi}_0d\sigma\exp{i\sigma(L_0-\bar{L}_0)}$, for an operator 
${\cal O}(w,\bar{w})$, we obtain
\eqn\pOp{
{\cal P}\,{\cal O}(w,\bar{w})\,{\cal P}=
\int^{2\pi}_0{d\sigma\o2\pi}~{\cal O}(we^{i\sigma},\bar{w}e^{-i\sigma})
{\cal P}.
}
By applying \pOp\ to \preVceps, we find that 
$\Vc{1^\rc,2^\rc,3^\rc}\ket{\delta\Phi}_3$ can be rewritten as 
the reflector $\bra{R^\rc(1^\rc,2^\rc)}$ operated by 
\eqn\Vceps{
\left({1\o\alpha'g^2}\right)\int^{2\pi}_0{d\sigma\o2\pi}
(c^{(1)}(\sigma)+\bar{c}^{(1)}(\sigma))\,\dB_{ij}\,
\d{X}^{i(1)}\,\bar\d{X}^{j(1)}(e^{i\sigma},e^{-i\sigma})\,
(b_0^-{\cal P})^{(1^\rc)},  
}
where $c(\sigma)=e^{-i\sigma}c(e^{i\sigma})=\sum_nc_ne^{-in\sigma}$ 
and $\bar{c}(\sigma)=e^{i\sigma}\bar{c}(e^{-i\sigma})
=\sum_n\bar{c}_ne^{in\sigma}$.
Furthermore, $\d{X}^i\bar\d{X}^j(e^{i\sigma},e^{-i\sigma})$ in \Vceps\ can 
be written in terms of the string coordinates and the momenta 
as $A^i(\sigma)\bar{A}^j(\sigma)$ where 
\eqn\Aop{\eqalign{
A^i(\sigma)&={1\o2i}g^{ij}\left[(2\pi\a')P_j(\sigma)
+\left(g_{jk}-b_{jk}\right){X'}^{k}(\sigma)\right],
\cr
\bar{A}^i(\sigma)&={1\o2i}g^{ij}\left[(2\pi\a')P_j(\sigma)
-\left(g_{jk}+b_{jk}\right){X'}^{k}(\sigma)\right].
\cr}
}
Substituting $A^i(\sigma)\bar{A}^j(\sigma)$ into \Vceps, we find that 
\Vceps\ becomes 
\eqn\preBRS{
{1\o g^2}
{2\pi\a'\o(2\alpha')}\int^{2\pi}_0{d\sigma\o2\pi}
(c^{(1)}(\sigma)+\bar{c}^{(1)}(\sigma))\,g^{ik}\dB_{kj}\,
\left(P_i(\sigma)-B_{il}{X'}^l(\sigma)\right){X'}^j(\sigma)\,
(b_0^-{\cal P})^{(1^\rc)}. 
}
As we can see from \delQB, \preBRS\ can certainly be interpreted as 
the change of the BRS charge and we obtain the final result of 
this calculation 
\eqn\cdelBRS{
\Vc{1^\rc,2^\rc,3^\rc}\ket{\delta\Phi}=-{1\o g^2}
\bra{R^\rc(1^\rc,2^\rc)}\delta\QB^{(1^\rc)}(B)(b_0^-{\cal P})^{(1^\rc)}, 
}
where we label the variation of the BRS charge $\QB(B)$ with the letter $B$ 
as $\delta\QB(B)$ to indicate its dependence on the background field $B_{ij}$. 
This result \cdelBRS\ is in agreement with the result in \KugoZwie.


The condensation of the $B$-field can also change the kinetic term of 
open string fields. The change comes from the contraction of the 
open-open-closed vertex $\bra{{U}_\Omega}$ 
and the shift $\ket{\delta\Phi}$. 
The corresponding LPP vertex $\uomg{1,2,3^\rc;\sigma_0}$ is specified by 
the Mandelstam mapping \HN
\eqn\Mandomg{\eqalign{
\rho&=\sum_{r=1,2,3,3^*}\alpha_r \ln(z-Z_r)
\cr
&=\alpha_r \ln w_r + \rho_0^{(r)},
}}
where the string length parameters $\alpha_r$ satisfy 
$\sum_r \alpha_r=0$, and $\alpha_3=\alpha_{3^*}$. 
When $\a_1,\a_3\geq0$ and $\a_2\leq0$, 
$\rho^{(2)}_0=\rho_0-i\sigma^{(1)}_0-i\pi\a_2$;
$\rho^{(3)}_0=\rho_0$; 
$\rho^{(1)}_0=\rho_0-i\sigma^{(1)}_0$ for 
$0\leq\sigma^{(1)}\leq(\sigma^{(1)}_0/\a_1)$ and 
$\rho^{(1)}_0=\rho_0-i\sigma^{(1)}_0+2i\pi\a_3$ for 
$(\sigma^{(1)}_0/\a_1)\leq\sigma^{(1)}\leq\pi$, where 
$\sigma^{(1)}={\rm Im}\ln w_1$. $\rho_0$ is the interaction point 
which is given by $\rho_0=\rho(z_0)$; $z_0$ is defined by 
$(\d\rho/\d z)(z=z_0)=0$. $\sigma^{(1)}_0$ is the moduli parameter of this 
vertex which runs from $0$ to $\pi\a_1$, which is the point where the second 
open string breaks into the first open string and the closed string. 
Henceforth, for simplicity, $\sigma^{(1)}_0$ will be denoted by $\sigma_0$.
The Koba-Nielsen variables $Z_1$, $Z_2$ of the open strings are real, 
while $Z_3$ and $Z_{3^*}$ of the closed string are complex numbers 
satisfying that $Z_{3^*}={Z_3}^*$. 

Before calculating the contraction $\Uomg{1,2,3^\rc}\ket{B^{ij}}_B$, 
it is useful to introduce the open string counterpart of the operator 
${\cal B}$, which will be denoted by ${\cal B}_o$: 
${\cal B}_o=-(i/2)\int^{\pi}_0d\sigma B_{ij}{X'}^i(\sigma)X^j(\sigma)$. 
Since the vertex $\Uomg{1,2,3^\rc}$ in the case where $\alpha_1,\alpha_3\geq0$ 
$\alpha_2\leq0$ satisfies the overlapping condition
\eqn\OmgOverLap{\eqalign{
&\uomg{1,2,3^\rc;\sigma_0}\left[{X}^{i(1)}\left({\sigma\o\alpha_1}\right)
-{X}^{i(2)}\left({\pi|\alpha_2|-\sigma\o|\alpha_2|}\right)\right]=0, 
\quad\left(0\leq\sigma<\sigma_0\right)
\cr
&\uomg{1,2,3^\rc;\sigma_0}\left[{X}^{i(3)}
\left({\sigma-\sigma_0\o\alpha_3}\right)
-{X}^{i(2)}\left({\pi|\alpha_2|-\sigma\o|\alpha_2|}\right)\right]=0, 
\quad\left(\sigma_0\leq\sigma<\sigma_0+2\pi\alpha_3\right)
\cr
&\uomg{1,2,3^\rc;\sigma_0}\left[{X}^{i(1)}
\left({\sigma-2\pi\alpha_3\o\alpha_1}\right)
-{X}^{i(2)}\left({\pi|\alpha_2|-\sigma\o|\alpha_2|}\right)\right]=0, 
\left(\sigma_0+2\pi\alpha_3\leq\sigma\leq\pi|\alpha_2|\right)
\cr}}
we can see that 
\eqn\OmgBBB{
\Uomg{1,2,3^\rc}\left[{\cal B}^{(1)}_o+{\cal B}^{(2)}_o+{\cal B}^{(3)}\right]
=0.
}
Since, by our assumption, the vertex $\Uomg{1,2,3^\rc}$ is background independent 
and $\ket{B^{ij}}_B=e^{\cal B}\ket{B^{ij}}_0$, we can see from \OmgBBB\ that 
\eqn\OmgBeBeB{
\Uomg{1,2,3^\rc}\ket{B^{ij}}_B
=\Uomg{1,2,3^\rc}\ket{B^{ij}}_0e^{-{\cal B}^{(1)}_o-{\cal B}^{(2)}_o}. 
}
Thus, in the following, we will calculate the contraction 
$\Uomg{1,2,3^\rc}\ket{B^{ij}}_0$, where there is no background $B$-field. 
Hence, for the sake of simplicity, we will omit from the states and the 
oscillators the label $0$, which means no background $B$-field. 
After the calculation, we will restore the label. 

In a similar way to what we have done for the closed three-string vertex, 
using the property of the LPP vertex, we find that 
\eqn\Uomgeps{\eqalign{
&\Uomg{1,2,3^\rc}\ket{\delta\Phi}_{3^\rc}
\ket{A}_2\ket{B}_1
\cr
&=\lim_{\alpha_3\rightarrow 0} -{2x_{\Omega}\o\a'x_cg^2}\delta B_{ij}
\int_{\sigma_i}^{\sigma_f} d\sigma_0
\left\langle b_{\sigma_0}\,c\bar{c}\d X^i \bar{\d} X^j(Z_3,{Z_3}^*)\,
h_2^{\epsilon}[{\cal O}_A]\,h_1^{\epsilon}[{\cal O}_B]
\right\rangle.
}}
Here the anti-ghost factor is given \AKTa\ by
\eqn\antighost{
b_{\sigma_0}={d\rho_0 \o d\sigma_0} \oint_{C_{\rho_0}}
{d\rho\o 2\pi i} b(\rho)
+{d\rho_0^* \o d\sigma_0} \oint_{C_{\rho_0^*}}
{d\rho^*\o 2\pi i} \bar{b}(\rho^*).
}
We can rewrite the anti-ghost factor as
\eqn\antighostZ{\eqalign{
b_{\sigma_0}&= i \oint_{C_{z_0}}
{dz\o 2\pi i} \left({d\rho\o dz}\right)^{-1}b(z)
-i \oint_{C_{z_0^*}}
{dz^*\o 2\pi i} \left({d\rho^*\o dz^*}\right)^{-1} \bar{b}(z^*)
\cr
&=2i \left({d^2\rho\o dz^2}\right)^{-1}_{z=z_0}b(z_0)
-2i \left({d^2\rho^*\o dz^{*2}}\right)^{-1}_{z^*=z^*_0}\bar{b}(z^*_0),
}}
where we have used 
\eqn\exprho{
\rho(z)=\rho(z_0)+{1\o 2}\left({d^2\rho\o dz^2}\right)^2_{z=z_0}(z-z_0)^2
+{\rm O}((z-z_0)^3).
}
As we can see from the Mandelstam mapping of \Mandomg, 
when $\a_3$ goes to zero, we can obtain the following equations: 
\eqn\intpoint{\eqalign{
z_0&=Z_3 -\alpha_3 {(Z_3-Z_1)(Z_3-Z_2)\o \alpha_1 Z_1+\alpha_2 Z_2}
+{\rm O}(\alpha_3^2),
\cr
\left({d^2\rho\o dz^2}\right)^{-1}_{z=z_0}
&=-\alpha_3 \left[
{(Z_3-Z_1)(Z_3-Z_2)\o \alpha_1 Z_1+\alpha_2 Z_2}
\right]^2 +{\rm O}(\alpha_3^2).
}}
The first equation means that the interaction point $z_0$ goes to $Z_3$ in 
the limit $\a_3\rightarrow0$.
Then, we can evaluate the ghost part of the correlation function \Uomgeps,
\eqn\UomgepsGhost{\eqalign{
\lim_{\alpha_3\rightarrow 0}
b_{\sigma_0} c(Z_3)\bar{c}(Z_3^*)
&=
\lim_{\alpha_3\rightarrow 0}\,2i \left[
\left({d^2\rho\o dz^2}\right)^{-1}_{z=z_0}
{1\o z_0-Z_3}\bar{c}(Z_3)
+ \left({d^2\rho^*\o dz^{*2}}\right)^{-1}_{z^*=z_0^*}
{1\o z_0^*-Z_3^*}c(Z_3)
\right]
\cr
&=
2i\left[{(Z_3-Z_1)(Z_3-Z_2)\o \alpha_1 Z_1+\alpha_2 Z_2}\bar{c}(Z_3)
+{(Z_3^*-Z_1)(Z_3^*-Z_2)\o \alpha_1 Z_1+\alpha_2 Z_2}
c(Z_3)
\right]
\cr
&=2i\left[\left({d\tilde{\rho}\o dz}\right)^{-1}_{z=Z_3}
\bar{c}(Z_3^*)
+\left({d\tilde{\rho}^*\o dz^*}\right)^{-1}_{z^*=Z_3^*}
c(Z_3)\right],
}}
where $\tilde{\rho}$ denotes the mapping \Mandomg\ in the limit 
$\alpha_3\rightarrow 0$. In the same limit, since $z_0\rightarrow{Z_3}$, 
the insertion point of the operator $c\bar{c}\d{X}^i\bar{\d}{X}^j$ becomes 
equal to the interaction point $\rho_0$ on the $\rho$ plane. 
Because this point corresponds to the point $\sigma^{(1)}=(\sigma_0/\a_1)$ 
on the worldsheet of the first string, we can see that 
\eqn\confmap{\eqalign{
&\left({d\tilde{\rho}\o dz}\right)^{-1}
\bar{c}\d X^i \bar{\d} X^j(z_0,{z_0}^*)
={e^{i(\sigma_0/\a_1)}\o\a_1}h_1\left[\bar{c}
\d X^i \bar{\d} X^j(e^{i(\sigma_0/\a_1)},e^{-i(\sigma_0/\a_1)})\right],
\cr
&\left({d\tilde{\rho}^*\o dz^*}\right)^{-1}
c\d X^i \bar{\d} X^j(z_0,{z_0}^*)
={e^{-i(\sigma_0/\a_1)}\o\a_1}h_1\left[{c}
\d X^i \bar{\d} X^j(e^{i(\sigma_0/\a_1)},e^{-i(\sigma_0/\a_1)})\right],
}
}
where we have used $(dz/d\tilde{\rho})=(w_1/\a_1)(dz/dw_1)$ and the 
anti-holomorphic counterpart, when the both sides are contracted with 
$\delta{B}_{ij}$.

Substituting \UomgepsGhost\ and \confmap\ into \Uomgeps\ and scaling $\sigma_0$ 
as $\sigma_0\rightarrow\alpha_1\sigma_0$, we find that 
$\Uomg{1,2,3^\rc}\ket{\delta\Phi}_3$$\ket{A}_2\ket{B}_1$ becomes 
\eqn\UomgPhi{
-\bra{R^o(1,2)}{8\pi ix_{\Omega}\o\a'x_cg^2}\delta B_{ij}
\int_0^\pi {d\sigma_0\o 2\pi}\,
(\rho_0^*c^{(1)}(\rho_0)+\rho_0\bar{c}^{(1)}(\rho_0^*))\,
\d X^{i(1)}\bar{\d} X^{j(1)}(\rho_0,\rho_0^*)\ket{A}_2\ket{B}_1,
}
where $\rho_0=e^{i\sigma_0}$. In the paper \AKTa, it was shown that 
the BRS invariance of the action requires that $8\pi{i}x_{\Omega}=x_c$, 
which we will apply to \UomgPhi.
Since $\d{X}^i(w)=(l_s/2i)\sum_n\a^i_nw^{-n-1}$ 
and $\bar\d{X}^i(w^*)=(l_s/2i)\sum_n\a^i_n(w^*)^{-n-1}$, 
in terms of the string coordinates 
$i\d_{\sigma}X^i(\sigma)=l_s\sum_{n\not=0}\a^i_n\sin(n\sigma)$ and the momenta 
$P_i(\sigma)=(1/\pi\l)\sum_ng_{ij}\a^j_n\cos(n\sigma)$ for open strings, 
the operator $\d{X}^i\bar\d{X}^j(\rho_0,\rho_0^*)$ turns out to be 
$-(1/4)\left(2\pi\a'g^{ik}P_k+{X'}^i\right)$ 
$\left(2\pi\a'g^{jl}P_l-{X'}^j\right)$. Therefore, we can verify that 
\UomgPhi\ becomes 
\eqn\UomgPhiSigma{
-\bra{R^o(1,2)}{1\o g^2}{2\pi\a'\o(2\alpha')}\int^{\pi}_0{d\sigma\o2\pi}
\left(c^{(1)}(\sigma)+\bar{c}^{(1)}(\sigma)\right)\,g^{ik}\dB_{kj}\,
P_i^{(1)}(\sigma){X'}^{j(1)}(\sigma)\ket{A}_2\ket{B}_1.
}
Here $\sigma_0$ has been renamed $\sigma$.
Thus, as we can see from \OmgBeBeB, 
$\Uomg{1,2,3^\rc}\ket{\delta\Phi}_{B\,3}$ in the presence of the $B$-field 
becomes 
\eqn\OmgPhi{
-\bra{R^o(1,2)}{1\o g^2}e^{{\cal B}_o^{(1)}}\int^{\pi}_0{d\sigma}
\hf\left(c^{(1)}(\sigma)+\bar{c}^{(1)}(\sigma)\right)\,g^{ik}\dB_{kj}\,
P_i^{(1)}(\sigma){X'}^{j(1)}(\sigma)e^{-{\cal B}_o^{(1)}}, 
}
where the operator $e^{-{\cal B}_o^{(2)}}$ have been passed through the 
operator acting on the reflector $\bra{R^o(1,2)}$ in \UomgPhiSigma\ and 
made into $e^{{\cal B}_o^{(1)}}$ by using the reflector. 

Now, let us show that the operator 
\eqn\delQBOpen{
e^{{\cal B}_o}\int^{\pi}_0{d\sigma}
\left(c(\sigma)+\bar{c}(\sigma)\right)\,g^{ik}\dB_{kj}\,
P_i(\sigma){X'}^{j}(\sigma)e^{-{\cal B}_o}
}
is equal to the variation of the BRS charge for the open strings.
We have 
\eqn\betaP{
\left[{\cal B}_o, P_i(\sigma)\right]=-\hf B_{ij}
\left[\pder{\sigma}\int^{\pi}_0d\sigma'\delta_s(\sigma,\sigma'){X}^j(\sigma')
+\int^{\pi}_0d\sigma'\delta_c(\sigma,\sigma'){X'}^j(\sigma')\right],
}
where we define that 
$\delta_s(\sigma,\sigma')=
(2/\pi)\sum^{\infty}_{n=1}\sin(n\sigma)\sin(n\sigma')$ 
and that 
$\delta_c(\sigma,\sigma')=(1/\pi)+(2/\pi)\sum^{\infty}_{n=1}
\cos(n\sigma)\cos(n\sigma')$. We can also prove that 
\eqn\delsinarrowdelcos{
\int^{\pi}_{0}d\sigma d\sigma' f(\sigma)\delta_c(\sigma,\sigma')g(\sigma')
=\int^{\pi}_{0}d\sigma d\sigma' f(\sigma)\delta_s(\sigma,\sigma')g(\sigma')
}
where $f(\sigma)$ and $g(\sigma)$ are $\cos(n\sigma)$ or $\sin(n\sigma)$ 
with $n\geq0$. Furthermore, even when either $f(\sigma)$ or $g(\sigma)$ is 
$\sigma$, \delsinarrowdelcos\ still holds. 
By using \betaP\ and \delsinarrowdelcos, 
we find that \delQBOpen\ becomes 
\eqn\delQBOpen{
\int^{\pi}_0{d\sigma}
\left(c(\sigma)+\bar{c}(\sigma)\right)\,g^{ik}\dB_{kj}\,
\left[P_i(\sigma)-B_{il}{X'}^{l}(\sigma)\right]{X'}^{j}(\sigma), 
}
which is equal to $\delta\QB(B)$; $\delta\QB(B)$ is 
the variation of the BRS charge $\QB(B)$ 
under $B_{ij}\rightarrow B_{ij}+\delta B_{ij}$. 
Here we assume that the BRS charge for open strings is 
defined as 
\eqn\oBRS{
\QB(B)=\int^{\pi}_0{d\sigma\o2\pi}\left[c(\sigma)T_{++}(\sigma)+
\bar{c}(\sigma)T_{--}(\sigma)\right], 
}
where the energy-momentum tensor ${T}_{\pm\pm}$ is 
\eqn\hTXP{
{1\o 4\alpha'}g^{ij}
\left[(2\pi\alpha')P_i\pm\left(g_{ik}\mp b_{ik}\right)
{X}^{\prime{k}}\right]
\left[(2\pi\alpha')P_j\pm\left(g_{jl}\mp b_{jl}\right)
{X}^{\prime{l}}\right]. 
}
Note that this BRS charge $\QB(B)$ is different from the BRS charge $\hat{Q}_B$ 
in the previous section in the usage of $X^i(\sigma)$ instead of 
$\hat{X}^i(\sigma)$. 
Thus, we obtain that 
\eqn\odelBRS{
\Uomg{1,2,3^\rc}\ket{\delta\Phi}_{B}=
-{1\o g^2}\bra{R^o(1,2)}\delta\QB^{(1)}(B).
}

At last, we can find another string field theory in the finite $B$-field by 
making use of \cdelBRS\ and \odelBRS: 
\eqn\commaction{\eqalign{
S(B)=-&{1\o 2}{\bra{\Psi}}\QB(B){\ket{\Psi}}
-{1\o 2}\bra{\Phi}\QB(B)(b_0^-{\cal P})\ket{\Phi}
\cr
&+{g\o 3}\Vo{1,2,3} {\ket{\Psi}}_{321}
+{g^2\o 4}\Vfo{1,2,3,4}{\ket{\Psi}}_{4321}
+{g^2\o 3!}\Vc{1^\rc,2^\rc,3^\rc}\ket{\Phi}_{321}
\cr
&+g \U{1,2^\rc}\ket{\Phi}_2{\ket{\Psi}}_1
+{g^2 \o 2}\Uomg{1,2,3^\rc}\ket{\Phi}_3{\ket{\Psi}}_{21}, 
}}
where $\QB(B)$ for closed strings is given by \cBRS\ and for open strings 
by \oBRS.


\newsec{The Unitary Transformation between the Two SFTs}

In the previous sections we have constructed two open-closed string field 
theories: one, given in section $2$, has the noncommutative factor 
$\exp[-\sum_{r<s}(i/2)\theta^{ij}p^{(r)}_ip^{(s)}_j]$ in the open three- and 
four-string vertices and the unitary operators $e^{M}$ put on the legs of open 
string fields in the other vertices, while the other, shown in section $3$, has 
the same vertices as the theory in the absence of the $B$-field. 
The string coordinates $\hX^i(\sigma)$ in the BRS charge $\hat\QB$ of the former 
theory is replaced with the background independent coordinates $X^i(\sigma)$ in 
the BRS charge $\QB$ of the latter theory. Thus, since we have two descriptions 
of the same physics, we are led to ask what relation exists between these 
theories. 

Henceforth, we will call the former theory in section $2$ `noncommutative' 
theory and the latter in section $3$ `commutative' theory.  As we have 
shown in section 2, in the noncommutative theory, each open string field has the 
unitary operator $e^{M}$. If the open string fields $\hat{\Psi}$ are transformed 
as 
\eqn\eMoSF{
\ket{\Psi}=e^{M}\hat{\ket{\Psi}},
}
all the interactions in the noncommutative theory 
become the same as those in the commutative theory. Then, the BRS charge 
$\hat\QB$ in the noncommutative theory becomes $e^{M}\hat\QB e^{-M}$. 
We will show below that the transformed BRS charge 
$e^{M}\hat\QB e^{-M}$ is equal to the BRS charge $\QB$ in the commutative 
theory. Thus, by the unitary transformation, these theories are related to 
each other. 

By the unitary transformation, the `noncommutative' string coordinates 
$\d_{\sigma}\hat{X}^i(\sigma)$ turn into 
\eqn\UX{
e^{M}\d_{\sigma}\hat{X}^i(\sigma)e^{-M}
=\d_{\sigma}X^i(\sigma)+\theta^{ij}\int^{\pi}_{0}d\sigma'
\left[\delta_c(\sigma,\sigma')-\delta_s(\sigma,\sigma')\right]P_j(\sigma').
}
The second term in the right-hand side of \UX\ can be seen 
from \delsinarrowdelcos\ to be vanishing, when it is included in an integrand. 
Therefore, in the transformed BRS charge $e^{M}\hat\QB e^{-M}$, all we have 
to do is to replace the string coordinates $\hat{X}^i(\sigma)$ with the ordinary 
coordinates $X^i(\sigma)$. This means that $e^{M}\hat\QB e^{-M}$ is equal to 
the BRS charge $\QB$ in the commutative theory. Thus, the noncommutative theory 
is mapped to the commutative theory by the unitary transformation \eMoSF.

\newsec{Discussion}

In this paper, we have obtained two open-closed string field theories in the 
same background. One gives a noncommutative gauge theory in the low-energy 
limit, while the other is expected to give an ordinary gauge theory with 
the constant background field strength $B_{ij}$ in the infrared. 
In section $4$, we have shown that these theories can be transformed into 
each other by the unitary operator. 

As we mentioned in the footnote in section $2$, our oriented open-closed 
string field theory does not maintain the full gauge invariance. 
Since we need worldsheets to be orientable to include the background $B$-field, 
it would be preferable to have an open-closed string field theory 
in an appropriate curved background to maintain the gauge invariance by using 
the Fischler-Susskind mechanism \FS, while the orientability of worldsheets is 
kept, as has been argued in \KugoSuppl. 
Besides this issue, the condensation of closed string fields we have 
dealt with only satisfies the equation of motion from the string 
field theory up to the next leading order of the string coupling constant $g$, 
mainly because, at least at present, we do not have the complete action of 
a fully gauge invariant oriented open-closed string field theory. 
However, everything we have shown in this paper would remain true at the 
leading order of the coupling constant $g$, even in 
such a gauge invariant string field theory. 

This paper was motivated by the papers \refs{\hiko,\Sugino}. 
Since the kinetic term in Witten's string field theory is the same as the 
kinetic term in our string field theory in this paper, it is natural to 
think that Witten's theory transformed by the unitary operator could give 
an ordinary gauge theory with the background field strength in the low-energy 
limit. Although, in Witten's open string field theory, we do not 
have explicit closed string fields, by using the above transformed theory, 
we might be able to discuss an infinitesimal variation of the background 
$B$-field in a similar way to the discussion in \Strominger. 
This may lead us to the understanding of the gauge symmetry 
$B_{ij}\rightarrow B_{ij}+\d_i\Lambda_j-\d_j\Lambda_i$, 
$A_i\rightarrow A_i+\Lambda_i$ in Witten's string field theory.


\medskip
\centerline{{\bf Acknowledgements}}
The authors would like to thank Tsuguhiko Asakawa, Taichiro~Kugo, Katsumi~Itoh, 
Masahiro~Maeno, Kazumi Okuyama, Kazuhiko Suehiro, and Seiji Terashima 
for valuable discussion. 
T.$\,$K. is grateful to 
the organizers and the participants of Summer Institute '99 in Yamanashi, Japan
for the hospitality and stimulating atmosphere created there, which helped 
to initiate this work. 
T.$\,$K. was supported in part by a Grant-in-Aid (\#11740143)
and in part by a Grant-in-Aid for Scientific Research 
in a Priority Area: ``Supersymmetry and Unified Theory of Elementary 
Particles''(\#707), from the Ministry of Education, Science, Sports and 
Culture.
T.$\,$T. was supported in part by Research Fellowships of the Japan Society for
the Promotion of Science for Young Scientists.

\appendix{A}{The Derivations of the Vertices in the Background $B$-Field}

In this appendix, we will derive the open-closed transition vertex 
$\hU{1,2^\rc}$ and the open-open-closed string vertex $\hUomg{1,2,3^\rc}$ 
in the presence of the background $B$-field. 

We begin with the open-closed transition vertex $\hU{1,2^\rc}$, which is 
specified up to an overall normalization by the overlapping condition 
\eqn\hucon{
\hU{2,1^\rc}\left\{
{\phi}^{(1^\rc)}(\sigma^{(1)})
-\hat{\phi}^{(2)}(\sigma^{(2)})
\right\}
=0, \quad (0\leq \sigma \leq 2\pi|\alpha_1|)
}
where $2\alpha_1+\alpha_2=0$ and $\sigma^{(1)}=\sigma/|\alpha_1|$; 
$\sigma^{(2)}=(\pi|\alpha_2|-\sigma)/|\alpha_2|$.
Here ${\phi}^{(1^\rc)}$ denotes $X^{i(1)}$ or $P^{(1)}_i$ of the first closed 
string and $\hat{\phi}^{(2)}$ denotes $\hX^{i(2)}$ or $P^{(2)}_i$ of the second 
open string. Similarly, the open-closed transition vertex $\U{1,2^\rc}$ in the 
absence of the $B$-field is specified by 
\eqn\ucon{
\U{2,1^\rc}\left\{
{\phi}^{(1^\rc)}(\sigma^{(1)})
-{\phi}^{(2)}(\sigma^{(2)})
\right\}
=0. \quad (0\leq \sigma \leq 2\pi|\alpha_1|)
}
Here, ${\phi}^{(2)}$ denotes $X^{i(2)}$ or $P^{(2)}_i$ of the second 
open string.

Recalling the definition of the operators $M$ for open strings from \M\ and 
for closed strings from \Mclose, by making use of \ucon\ we can verify that 
\eqn\uM{
\U{2,1^\rc}\left(M^{(1^\rc)}+M^{(2)}\right)=0.
}
We define the closed string counterpart of the `dual' coordinates $Q_j(\sigma)$ 
as 
$$
Q_j(\sigma)=\hf\int^{2\pi}_{0}d\sigma'\epsilon(\sigma-\sigma')P_j(\sigma'). 
$$
The dual coordinates $Q_j(\sigma)$ for closed strings connect to those for 
open strings on the open-closed transition vertex $\U{2,1^\rc}$ as
\eqn\Qcon{
\U{2,1^\rc}\left\{
Q_j^{(1^\rc)}(\sigma^{(1)})
-Q_j^{(2)}(\sigma^{(2)})
\right\}
=0. \quad (0\leq \sigma \leq 2\pi|\alpha_1|)
}
The commutation relation between $M$ and the string coordinates $X^i(\sigma)$ 
for closed strings turns out to be 
\eqn\closedMX{
\left[M, X^i(\sigma)\right]=-\theta^{ij}Q_j(\sigma).
}
From \closedMX, we find that
\eqn\UcXtoUoX{
\U{2,1^\rc}e^{M^{(1^\rc)}}X^{i(1^\rc)}(\sigma^{(1)})
=\U{2,1^\rc}e^{M^{(1^\rc)}}
\hX^{i(2)}(\sigma^{(2)}), 
\quad (0\leq \sigma \leq 2\pi|\alpha_1|)
}
and, thus, we obtain 
\eqn\UtU{
\hU{2,1^\rc}=\U{2,1^\rc} e^{M^{(2)}}.
}

Next we will move to the open-open-closed string interaction $\hUomg{3,2,1^\rc}$ 
in the presence of the $B$-field. The overlapping condition for the case 
$\alpha_1, \alpha_3\geq0$ and $\alpha_2\leq0$ is given by 
\eqn\hOmgOverLapp{\eqalign{
&\huomg{1,2,3^\rc;\sigma_0}\left[\hat{\phi}^{i(1)}(\sigma^{(1)})
-\hat{\phi}^{i(2)}(\sigma^{(2)})\right]=0, 
\quad\left(0\leq\sigma<\sigma_0\right)
\cr
&\huomg{1,2,3^\rc;\sigma_0}\left[{\phi}^{i(3^\rc)}(\sigma^{(3)})
-\hat{\phi}^{i(2)}(\sigma^{(2)})\right]=0, 
\quad\left(\sigma_0\leq\sigma<\sigma_0+2\pi\alpha_3\right)
\cr
&\huomg{1,2,3^\rc;\sigma_0}\left[\hat{\phi}^{i(1)}
\left(\sigma^{(1)}-2\pi{\alpha_3\o\alpha_1}\right)
-\hat{\phi}^{i(2)}(\sigma^{(2)})\right]=0, 
\left(\sigma_0+2\pi\alpha_3\leq\sigma\leq\pi|\alpha_2|\right)
\cr}}
where the notations for $\hat{\phi}^{i(r)}$ 
are the same as those 
for the vertex $\hU{2,1^\rc}$, except $\sigma^{(1)}=\sigma/\alpha_1$; 
$\sigma^{(2)}=(\pi|\alpha_2|-\sigma)/\alpha_2$; 
$\sigma^{(3)}=(\sigma-\sigma_0)/\alpha_3$.
The overlapping condition for the vertex $\Uomg{3,2,1^\rc}$ in the absence of 
the $B$-field is given by 
\eqn\OmgOverLapp{\eqalign{
&\uomg{1,2,3^\rc;\sigma_0}\left[{\phi}^{i(1)}(\sigma^{(1)})
-{\phi}^{i(2)}(\sigma^{(2)})\right]=0, 
\quad\left(0\leq\sigma<\sigma_0\right)
\cr
&\uomg{1,2,3^\rc;\sigma_0}\left[{\phi}^{i(3^\rc)}(\sigma^{(3)})
-{\phi}^{i(2)}(\sigma^{(2)})\right]=0, 
\quad\left(\sigma_0\leq\sigma<\sigma_0+2\pi\alpha_3\right)
\cr
&\uomg{1,2,3^\rc;\sigma_0}\left[{\phi}^{i(1)}
(\sigma^{(1)}-2\pi{\alpha_3\o\alpha_1})
-{\phi}^{i(2)}(\sigma^{(2)})\right]=0. 
\left(\sigma_0+2\pi\alpha_3\leq\sigma\leq\pi|\alpha_2|\right)
\cr}}
As we can see from \OmgOverLapp, on the vertex $\hUomg{1,2,3^\rc}$, we find that 
\eqn\MMMPQ{
\Uomg{1,2,3^\rc}\left\{M^{(1)}+M^{(2)}+M^{(3)}\right\}=\Uomg{1,2,3^\rc}
i\theta^{ij}p^{(3)}_iQ^{(1)}_j\left({\sigma_0\o|\alpha_1|}\right). 
}
In addition, we can verify that 
\eqn\QOverLap{\eqalign{
&\uomg{\sigma_0}\left[{Q}^{(1)}_j(\sigma^{(1)})
-{Q}^{(2)}_j(\sigma^{(2)})\right]
=\uomg{\sigma_0}\hf p^{(3)}, 
\quad\left(0\leq\sigma<\sigma_0\right)
\cr
&\uomg{\sigma_0}\left[{Q}^{(3^\rc)}_j(\sigma^{(3)})
-{Q}^{(2)}_j(\sigma^{(2)})\right]
=-\uomg{\sigma_0}Q^{(1)}_j\left({\sigma_0\o|\alpha_1|}\right), 
\quad\left(\sigma_0\leq\sigma<\sigma_0+2\pi\alpha_3\right)
\cr
&\uomg{\sigma_0}\left[{Q}^{(1)}_j(\sigma^{(1)}-2\pi{\alpha_3\o\alpha_1})
-{Q}^{(2)}_j(\sigma^{(2)})\right]
=-\uomg{\sigma_0}\hf p^{(3)}_j,  
\left(\sigma_0+2\pi\alpha_3\leq\sigma\leq\pi|\alpha_2|\right)
\cr}}
where $\uomg{\sigma_0}$ is short for $\uomg{1,2,3^\rc;\sigma_0}$.
By utilizing \MMMPQ\ and \QOverLap, we can find that 
\eqn\hOmgtoOmg{
\hUomg{1,2,3^\rc}=\Uomg{1,2,3^\rc}e^{M^{(1)}}e^{M^{(2)}}.
}

\listrefs

\end